# Magnetic field dependent variable range hopping behaviour in resistivity at low temperatures in polycrystalline colossal magnetoresistive manganites: evidence of spin polarised tunnelling

P. Raychaudhuri† , P. Taneja, S. Sarkar, A. K. Nigam, P. Ayyub, R. Pinto

***Department of Condensed Matter and Materials Science,***

***Tata Institute of Fundamental Research,***

***Homi Bhabha Road, Colaba, Mumbai-400005,India.***

***Abstract:*** It has been observed that the low temperature magnetoresistance behaviour in polycrystalline colossal magnetoresistive (CMR) manganites differ significantly from the single crystals. The polycrystalline samples show large magnetoresistance at temperatures much below the ferromagnetic transition temperature where the magnetoresistance of single crystals is very small. This has conventionally been attributed to spin polarised tunnelling at the grain boundaries in polycrystalline samples. In this paper we show the existence of a variable range hopping behaviour in resistivity at low temperatures in polycrystalline CMR samples. This behaviour gets gradually suppressed under the application of magnetic field. We discuss the significance of these results with respect to spin polarised tunnelling.

†e-mail:pratap@tifrc4.tifr.res.in

Recent studies on colossal magnetoresistive (CMR) manganites has revived interest in spin polarised intergranular tunnelling in the polycrystalline form of these [1,2,3] and other oxide ferromagnets [4,5]. It has been observed that in the ferromagnetic phase many of these materials show much larger magnetoresistance in polycrystalline form than in the single crystals. This behaviour was first observed in perovskite CMR manganites [1,6,7] and later in reported in other ferromagnetic oxides, namely, chromium(IV) oxide [4] ($CrO_2$) and magnetite [5] ($Fe_3O_4$). The idea of spin polarised tunnelling in all these polycrystalline materials is as follows. The grains of these materials are separated by an insulating grain boundary through which the electron can tunnel conserving its spin angular momentum. Since the magnetisation is aligned in different direction in the two grains (due to different direction of their easy axes), they have different majority spin channels in the conduction band. Thus an electron suffers a spin dependent scattering when tunnelling from one grain to the other. The application of magnetic field aligns the magnetisation in the direction of the field thus reducing this scattering.

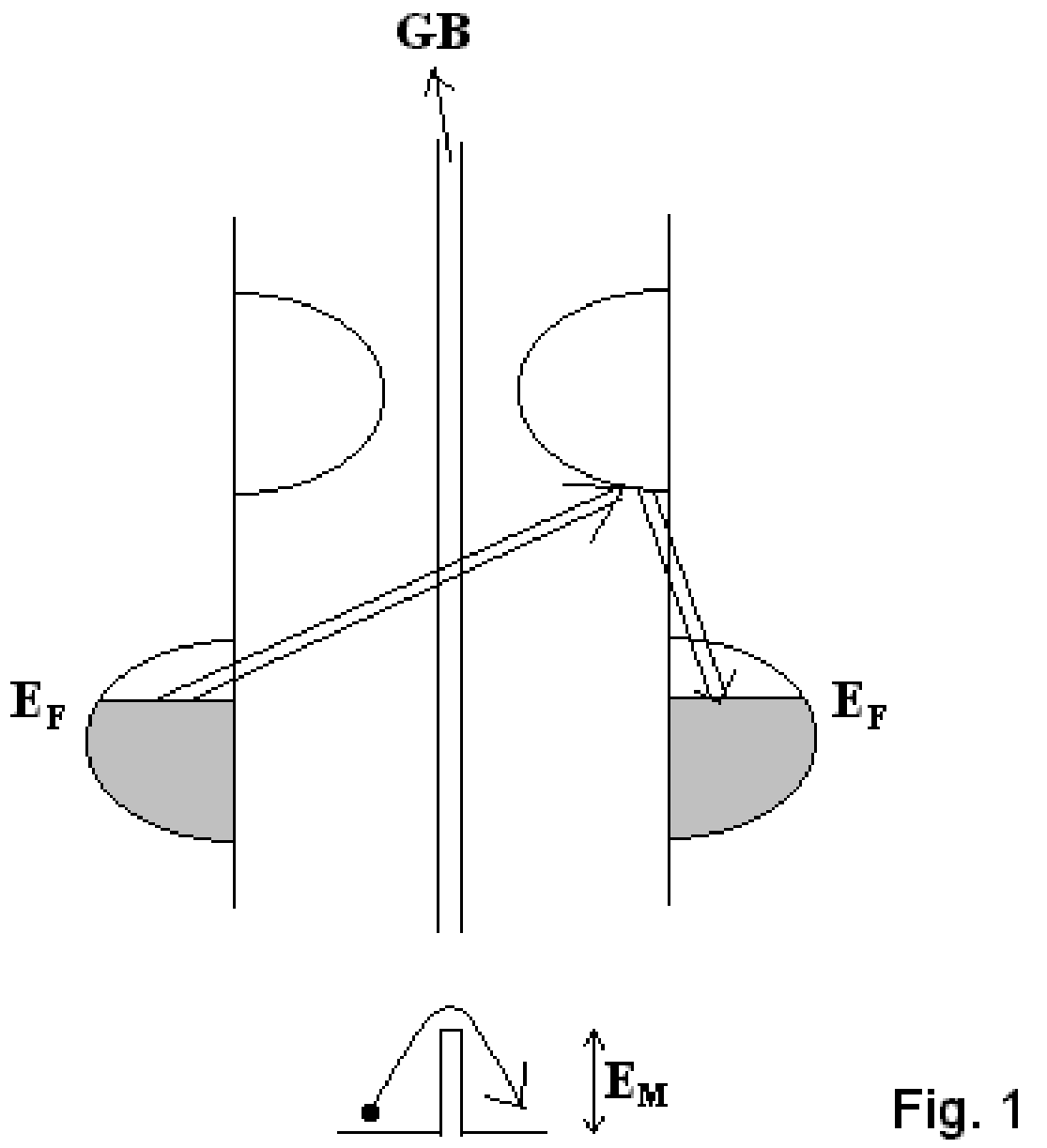

Fig. 1

Schematic representation of an electron passing across a grain boundary with the two grains having antiparallel magnetisation in the scenario of complete polarisation of the conduction band. An up (or down) spin from the Fermi level ($E_F$) of a grain initially goes to up (or down) subband of the other grain which is of higher energy, and then comes down to the majority spin channel via some dissipative process. This gives rise to an effective potential barrier $E_M$ as shown in the figure. The application of a magnetic field aligns the two grain magnetisation thus reducing this barrier.

In the case of CMR manganites the transport properties are understood in terms of Zener double exchange [8] mechanism which considers the magnetic coupling between $Mn^{3+}$ and $Mn^{4+}$ via the motion of an electron between two partially filled ***d*** shells with strong on site Hund's rule coupling. This results in complete polarisation of the $\boldsymbol{e_g}$ conduction band. In the double exchange picture the magnetoresistance should be simply related to the reduction in spin fluctuation. Thus one would ideally expect the magnetoresistance ( $MR \sim ((\rho(H)-\rho(0))/\rho(0) \sim \Delta\rho/\rho_0$ ) to monotonically drop below the ferromagnetic transition temperature ($T_c$) and become very small at temperatures much below $T_c$. However in polycrystalline CMR samples it has been seen that the magnetoresistance shows an increase with decreasing temperatures at low temperatures [1,9,10]. Also temperatures below $T_c$ the magnetoresistance in polycrystalline samples shows a sharp drop at low fields followed by a gradual smaller decrease. These features have been conventionally attributed to spin polarised tunnelling at the grain boundaries though non-tunnelling mechanisms have also been proposed to explain these features[7]. By separating out the intergranular and intragranular magnetoresistance we have earlier shown that the intergranular magnetoresistance drops off monotonically with increasing temperature whereas the intragranular part follows the behaviour expected from Zener double exchange[10]. In the presence of complete polarisation of the conduction band the electron will face a magnetic potential barrier while tunnelling between two grain with antiparallel magnetisation due to difference in energies of the respective up and down spin channels. In addition there could be a potential barrier coming from the insulating grain boundaries. The application of a magnetic field will align the grain magnetisations thus reducing the magnetic potential barrier. This is schematically shown in figure 1. This was the essence of the model of spin polarised tunnelling given by Hellman and Abeles [11]. Thus in the scenario of a is fully polarised conduction band one expects that the resistivity will show an increase with decreasing temperatures due to thermally activated or variable range hopping (VRH) conduction in the disordered barrier region at low temperature where the grain contribution to the resistivity is small [7] . In this paper we show the existence of such an increase at low temperatures in CMR manganites of different compositions which confirms the existence of spin polarised tunnelling in these materials. This increase fits to a VRH kind of behaviour. We also show that this increase is absent in epitaxial thin films where grain boundary effect is less dominant.

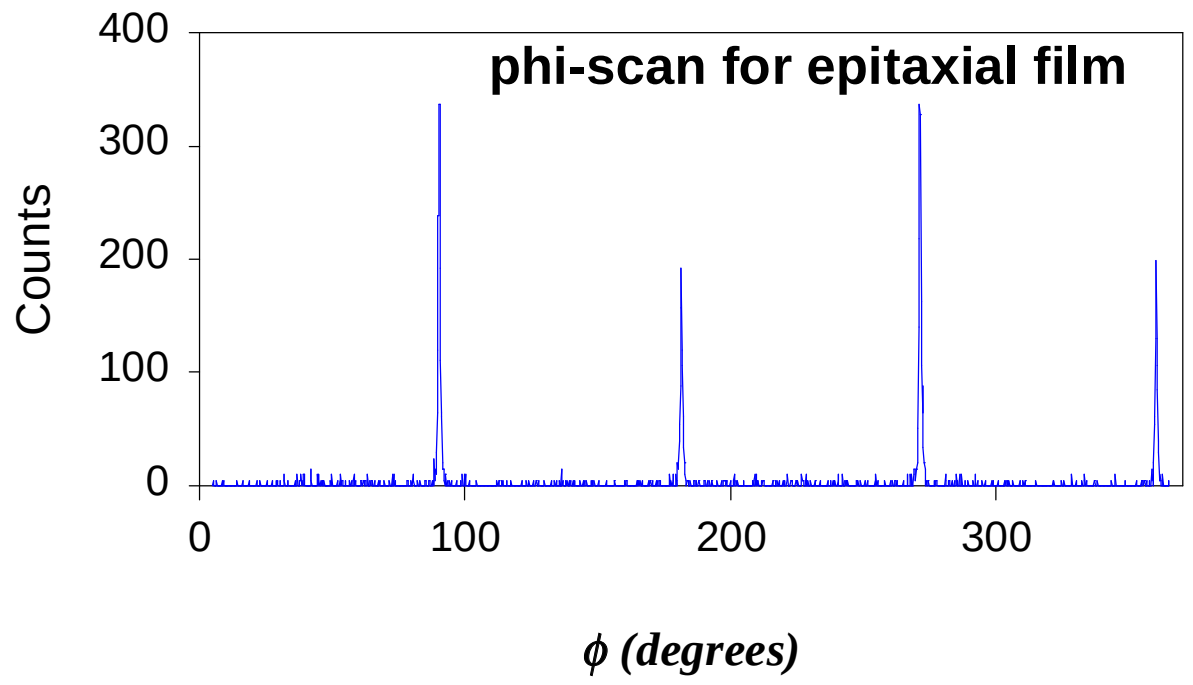

**Fig 2**

X-ray $\phi$ scan of the {101} family of peaks for the $La_{0.55}Ho_{0.15}Sr_{0.3}MnO_3$ film grown on $LaAlO_3$ showing very good in plane orientation.

Polycrystalline samples of $La_{0.7}Sr_{0.3}MnO_3$ ($T_c$~360 K) studied in this work was prepared by forming a stoichiometric mixture of nitrates of La, Mn and Sr and subsequently making the carbonates by adding ammonium carbonate solution. The precursors made of these carbonates were heated at $1200^0$ C for 24 hrs to get single phase $La_{0.7}Sr_{0.3}MnO_3$. $La_{0.55}Ho_{0.15}Sr_{0.3}MnO_3$ ($T_c$~250 K) and $La_{0.7}Sr_{0.3}Mn_{0.9}Ru_{0.1}O_3$ ($T_c$~350 K) samples were prepared through conventional solid sate reaction route starting from oxides and carbonates of the parent materials. These three polycrystalline samples were chosen since they have different $T_c$s and magnetic behaviour. All the samples were sintered between $1200^0$C and $1400^0$C. X-ray diffraction measurements confirmed all the samples to be single phase. Epitaxial and polycrystalline films of $La_{0.55}Ho_{0.15}Sr_{0.3}MnO_3$ were prepared by laser ablation on $LaAlO_3$ and polycrystalline yttria stabilised zirconia (YSZ) substrates respectively.

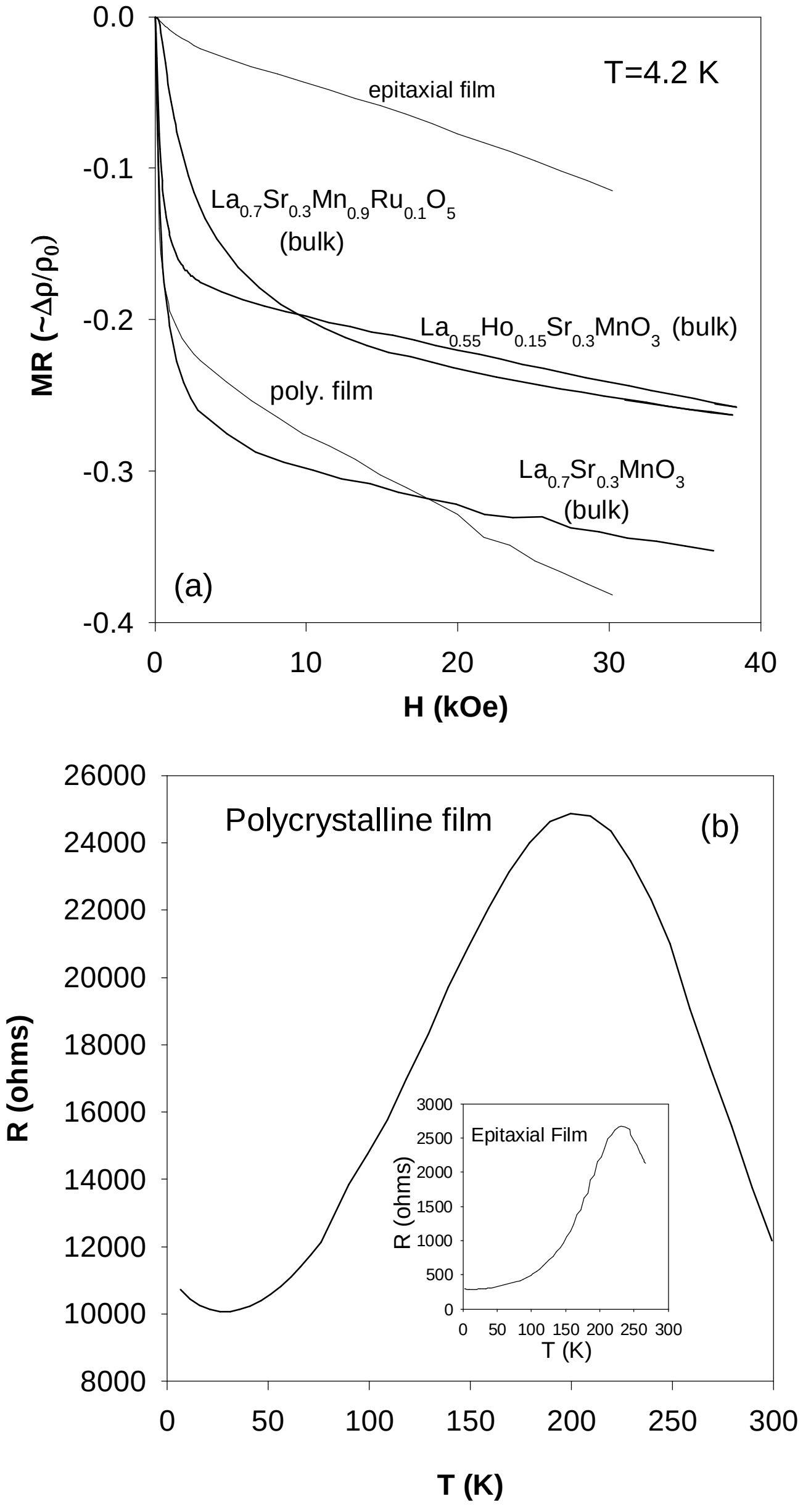

Fig. 3

(a) Magnetoresistance of various polycrystalline bulk and polycrystalline and epitaxial thin film at 4.2 K. All the polycrystalline bulk and thin film samples show a sharp drop at low fields arising due to spin polarised tunnelling at the grain boundaries. This sharp drop is absent for the epitaxial thin film. (b) Resistance versus temperature for the polycrystalline and epitaxial *(inset)* thin film. The polycrystalline film shows a clear increase in the resistance at low temperatures. This increase is absent for the polycrystalline film.

X-ray *θ–2θ* and *ϕ* scan measurements performed for the film grown on $LaAlO_3$ on a Philips four circle x-ray diffractometer confirmed the crystalline orientation and in plane alignment of the film. The off axis x-ray *ϕ* scan spectrum of the {101} family of peaks is shown in figure 2. The magnetoresistance was measured with conventional four probe technique using a home made superconducting magnet.

Figure 3a shows the MR as a function of field at 4.2 K for the thin films and polycrystalline bulk samples. All the polycrystalline samples show a sharp drop at low fields and have significant magnetoresistance at this temperature. This is associated with the magnetic domain wall scattering at grain boundaries [1,2,10]. The epitaxial film on the other hand does not show this feature. We thus observe that the epitaxial film behaves like a single crystal as reported earlier [2]. Figure 3b shows the resistance as a function of temperature for polycrystalline and epitaxial film (inset). The polycrystalline film clearly shows a rise in resistance at low temperatures which is for the epitaxial film. This rise in the resistance is a characteristic of polycrystalline manganite sample as we shall see in the next paragraph.

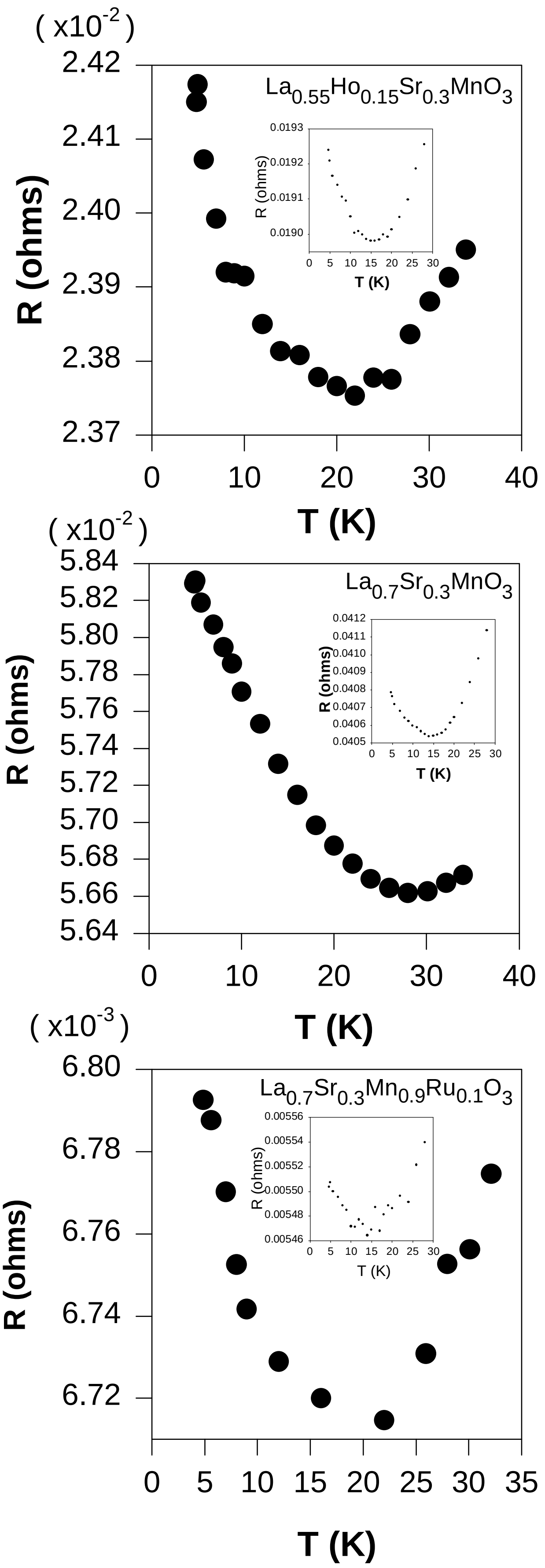

Fig 4

Resistance versus temperature (R-T) at zero field for the three polycrystalline bulk samples at low temperatures. There is an increase in resistance at low temperature for all these polycrystalline samples. The insets show the (R-T) curves in a field of 10 kOe. In the presence of field the resistance minimum has shifted to lower temperatures and the relative increase with respect to the minimum has decreased.

Figure 4 a-c show the resistance as a function of temperature (R-T) at low temperatures for the samples $La_{0.7}Sr_{0.3}MnO_3$, $La_{0.55}Ho_{0.15}Sr_{0.3}MnO_3$ and $La_{0.7}Sr_{0.3}Mn_{0.9}Ru_{0.1}O_3$ respectively. We observe a minimum in the resistance around 20 K in all these polycrystalline samples though their high temperature transport and magnetic properties are quite different. The insets show the R-T curves in an applied field of 10 kOe. We observe that the relative increase with respect to the minimum is largely suppressed at in an applied field. In figure 5 we show the R-T curves at low temperatures for $La_{0.7}Sr_{0.3}MnO_3$ in different applied fields. The resistance values are normalised to the value at the minimum ($R_{min}$). The interesting features are, i) the relative increase at low temperatures decreases with increasing fields and ii)the minima shift towards lower temperatures with increasing fields. The first feature can be understood from the fact that the application of a magnetic field suppresses the magnetic potential barrier at the grain boundary. The second feature can be realised from the fact that the resistance is composed of the grain contribution which decreases with field plus the grain boundary contribution which shows an increase. The grain contribution is similar to the single crystal and is relatively weakly dependent on field at low temperatures (the high field linear slope in the MR-H curve in figure 3a). Thus the suppression of the grain boundary contribution shifts the minimum to lower temperatures. We could fit the low temperature increase in the resistance to Mott's VRH expression without electron-electron interaction $R=R_\infty \exp((T_0/T)^{1/4})$ (inset of figure 6). Attempt to fit the expression to a thermally activated behaviour $R=R_0\exp(E_g/kT)$ gave a much poorer fit over the same temperature range. Figure 6 shows the values of $T_0$ as a function of field up to 35 kOe calculated by fitting the $R/R_{min}$-T data at various fields (figure 5) to the VRH formula. Though the absolute values of $T_0$ contain some errors arising due to the small range over which the data is fitted we clearly observed that $T_0$ decreases with rapidly with increasing field and becomes very small for field values greater than 10 kOe. This confirms the existence of a disordered magnetic potential barrier at the crystallographically disordered grain boundaries due to misaligned domains in the two grain. The application of a magnetic field aligns the grain magnetisations parallel to the applied field thus reducing this barrier. However the significance of the very small values of $T_0$ in this temperature range is not understood at this stage.

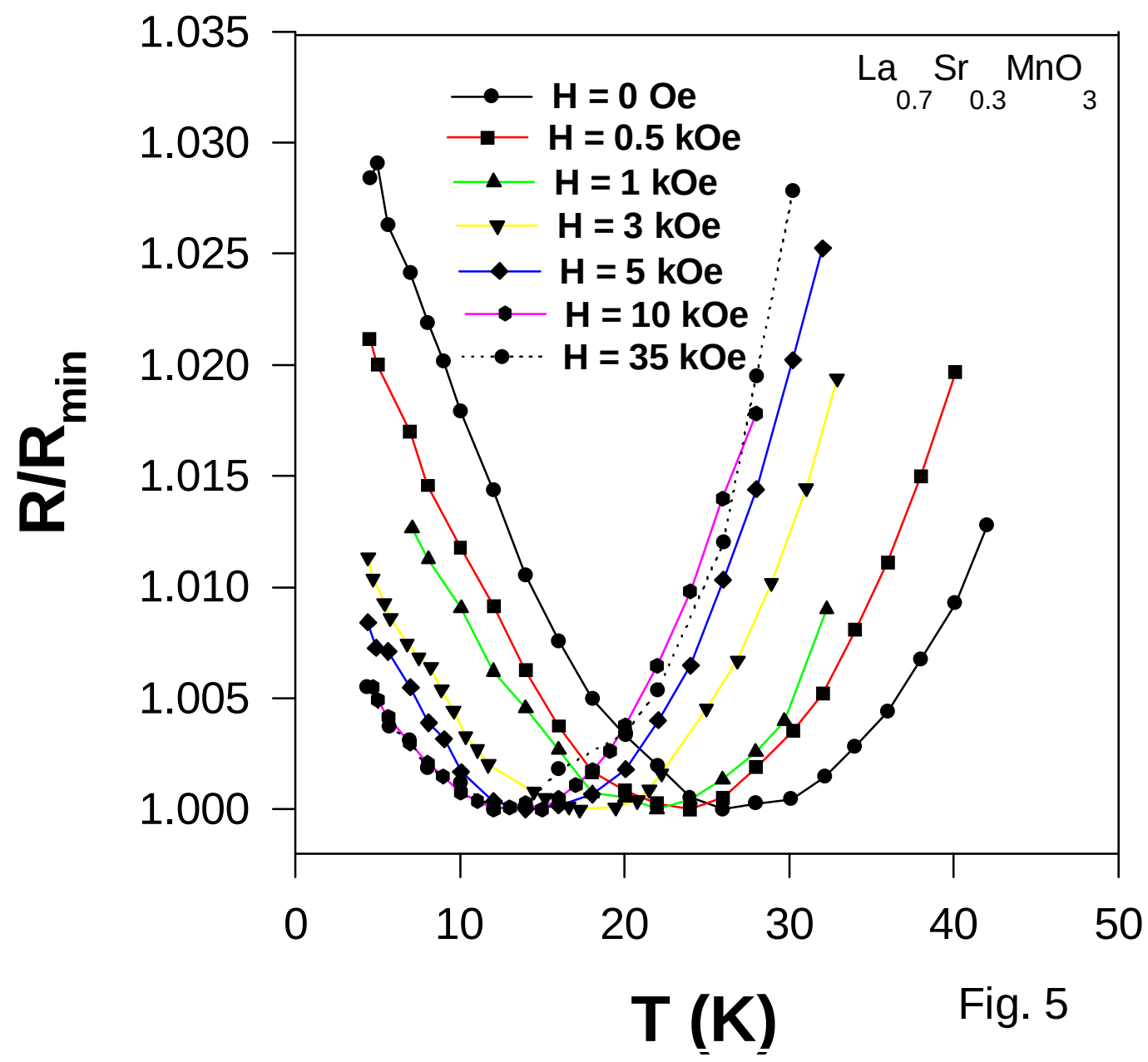

Fig. 5

The resistance (R) normalised to the value at the minimum ($R_{min}$) as a function of temperature at low temperatures (T) for $La_{0.7}Sr_{0.3}MnO_3$ polycrystalline sample at various fields.

These results show that at low temperature the resistance in polycrystalline CMR materials follow a VRH kind of behaviour. The scenario is similar to the one visualised in the model proposed by Hellman and Abeles [11]. However Hellman and Abeles assumed the transport to have an activated behaviour at the grain boundaries. In our case the good fit of the low temperature resistance obtained with the VRH behaviour suggests that the transport at the disordered grain boundary is of the VRH kind. This is significant due to the fact the resistivity in the disordered paramagnetic phase in these materials are known to follow the VRH kind of behaviour [12]. Observation of the same behaviour for the static spin disordered grain boundary region could be of interest in the context of manganite spin glasses [13,14,15] where a frozen static spin disorder exist. In this context it should also be noted that Hellman and Abeles' model was initially developed for granular nickel films where unlike CMR materials the conduction band is only 11% polarised. In the case of incomplete polarisation of the conduction band the tunnelling probability of an electron across two grains with antiparallel magnetisation should be decided by the respective up and down density of states in the two grains. It is not clear why in that scenario the electron will see a potential barrier when tunnelling between the two grains. In the case of CMR materials the low temperature VRH behaviour suggest that the model might be more appropriate in this context (at least at low temperatures).

In summary, we have observed a low temperature increase in the resistance at low temperatures in polycrystalline CMR samples. This increase is attributed to variable range hopping of the electron in the disordered grain boundary region. The variable range hopping parameter $T_0$ decreases by orders of magnitude under the application of magnetic field and almost saturates above 10 kOe suggesting that spin polarised tunnelling plays a dominant role in the intergranular transport of these materials. Our results thus provide further evidence for spin polarised tunnelling which has been invoked to explain the large low field magnetoresistance in polycrystalline CMR materials at temperatures below $T_c$.

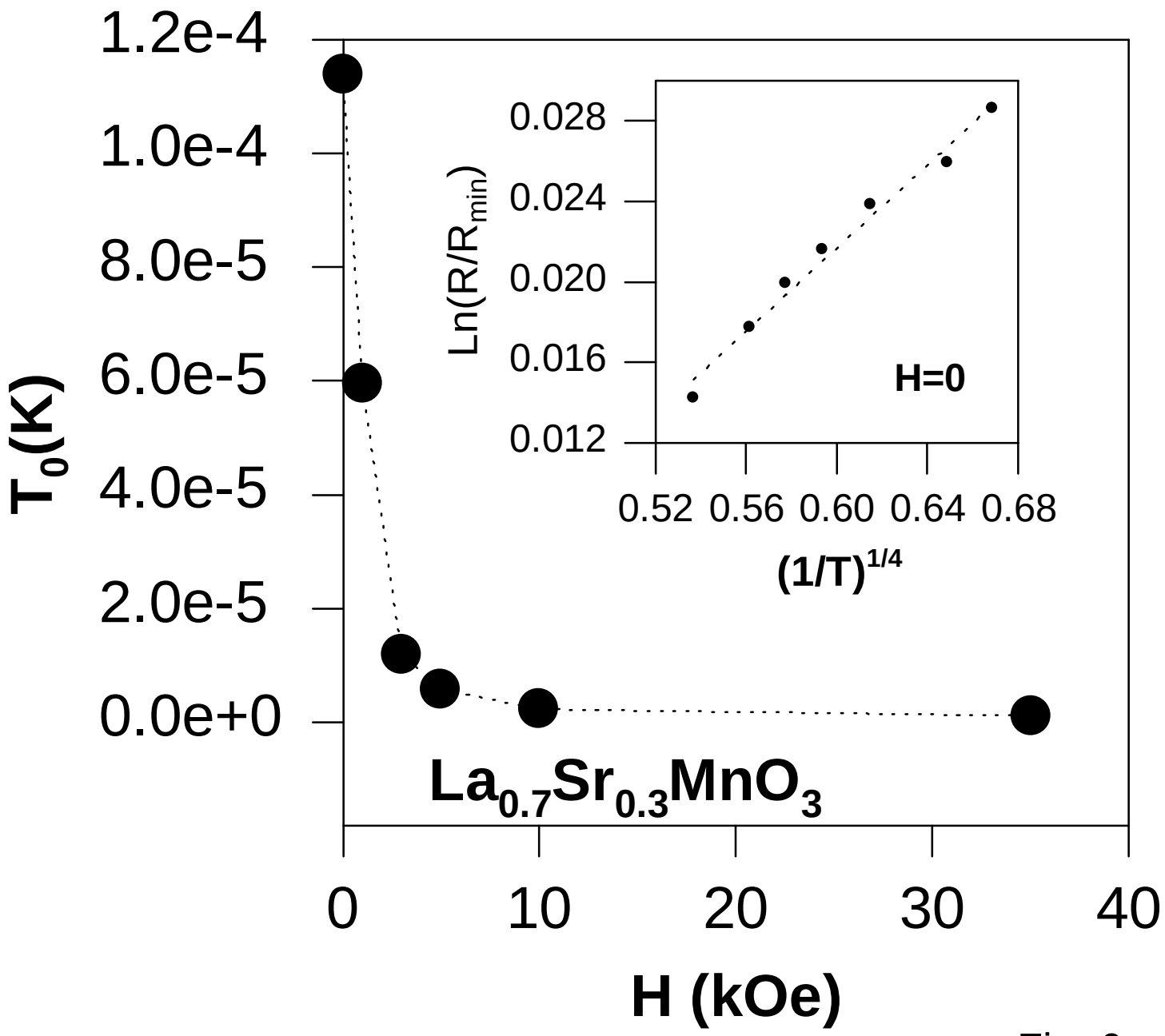

Fig. 6

The values of $T_0$ as a function of field deduced from the $R/R_{min}$ versus T curves using the VRH expression $R=R_\infty \exp((T_0/T)^{1/4})$ . The value of $T_0$ changes by orders of magnitude at an applied field of 10 kOe.

*Acknowledgements:* It is a pleasure to thank S Ramakrishnan for his suggestions and his keen interest in this work. We also thank R Sannabhadti and Arun Patade for technical help.